\documentclass[twocolumn,showpacs,amsmath,amssymb]{revtex4}
\usepackage{graphicx}

\begin{document}

\title{Comment on "Snapshot spectrum and critical phenomenon for two-dimensional classical spin systems"}

\author{Hiroaki Matsueda${}^{a}$}
\author{Ching Hua Lee${}^{b}$}
\author{Yoichiro Hashizume${}^{c}$}
\affiliation{
${}^{a}$Sendai National College of Technology, Sendai 989-3128, Japan \\
${}^{b}$Department of Physics, Stanford University, CA 94305, USA \\
${}^{c}$Department of Applied Physics, Tokyo University of Science, Tokyo 125-8585, Japan
}

\date{\today}
\maketitle

In Ref.~\cite{Matsueda}, one of the authors (HM) has examined scaling relations of snapshot entropy of classical spin configurations at criticality. The conclusion was that the snapshot entropy is a holographic entanglement entropy of one-dimensional near-critical systems. This argument is strongly supported by exact analysis of the entropy of fractal images and their dual one-dimensional quantum systems~\cite{Matsueda2}. Recently, more precise numerical examination for the spin snapshots has been done by Imura et. al~\cite{Imura}. They claim the presence of the anomalous dimension in the entropy scaling that could not be found in Ref.~\cite{Matsueda}. We agree with this claim, since the partial density matrix of the spin snapshot is roughly a two-point spin correlation function and we would observe the anomalous dimension of the scaling operator. Their interpretation is based on power law decay of the distribution function of the singular value spectrum, which is made by avaraging over many snapshots data, but the HM's original analysis was based on single snapshot. It seems to be quite useful to derive the Imura's result from viewpoints of the single snapshot. This is the purpose of this short note.

We briefly explain the definition of the snapshot entropy and outline of Imura's analysis. We start with the Ising model on the square lattice $H=-J\sum_{\left<i,j\right>}\sigma_{i}\sigma_{j}$, where $\sigma_{i}=\pm 1$ and the sum runs over the nearest neighbor lattice sites $\left<i,j\right>$, and $J (>0)$ is exchange interaction. The system size is taken to be $L\times L$. We denote the spin snapshot as $M(x,y)$, and regard it as a $L\times L$ matrix. The snapshot can be obtained by the Monte Carlo simulation. A key quantity is the reduced density matrix $\rho=MM^{\dagger}$ summing over $y$ degrees of freedom. If we denote the normalized eigenvalues of $\rho$ as $\lambda_{n}$ ($n=1,2,...,L$), the snapshot entropy is defined by
\begin{eqnarray}
S_{\chi} = -\sum_{n=1}^{\chi\le L}\lambda_{n}\ln\lambda_{n}.
\end{eqnarray}
In Ref.~\cite{Imura}, Imura has found the following scaling relation
\begin{eqnarray}
S_{\chi}=b\chi^{\eta}\ln\frac{\chi}{a}, \label{scaling}
\end{eqnarray}
and the new finding is the presence of the power term $\chi^{\eta}$ with $\eta=1/4$ for the Ising model. Actually, when we look at Fig. 6(a) in Ref.~\cite{Matsueda}, we observe some power factor as well as logarithmic term. This argument was based on idendification $\rho(x,x^{\prime})\sim\left<\sigma_{x,0}\sigma_{x^{\prime},0}\right>\propto \left|x-x^{\prime}\right|^{-\eta}$ for $\left|x-x^{\prime}\right|\gg 1$. If this identification is correct, the distribution function is given by
\begin{eqnarray}
f(\lambda)=\sum_{n}\delta\left(\lambda-\lambda_{n}\right)=A\lambda^{-\alpha}, \label{distribution}
\end{eqnarray}
with the exponent
\begin{eqnarray}
\alpha=\frac{2-\eta}{1-\eta}.
\end{eqnarray}
Then caluculating $S_{\chi}=\int_{\tilde{\lambda}}^{\infty}d\lambda f(\lambda)\lambda\ln\lambda$ with $\tilde{\lambda}\propto\chi^{\eta-1}$, we obtain Eq.~(\ref{scaling}).

\begin{figure}[htbp]
\begin{center}
\includegraphics[width=8cm]{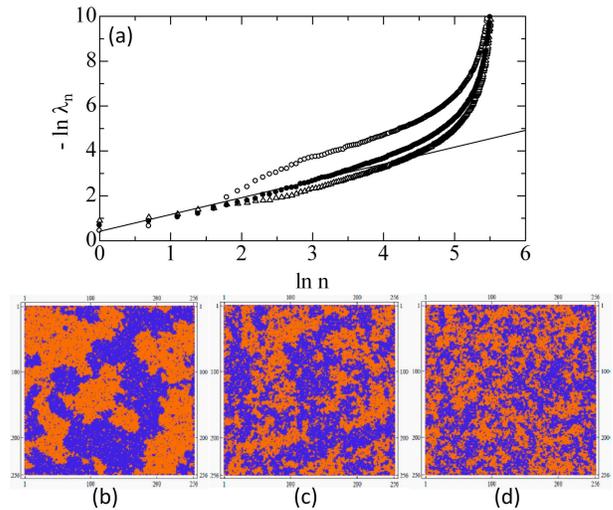}
\end{center}
\caption{(a) Singular value spectra $-\ln\lambda_{n}$ of snapshots with $L=256$ at $T=2.1J$ (open circles), $T=2.35J$ (filled circles), and $T=2.5J$ (open triangles). A fine solid line is a guide and the gradient is $1-\eta=0.75$. (b) Snapshot at $T=2.1J$. (c) Snapshot at $T=2.35J$. (d) Snapshot at $T=2.5J$.}
\label{fig1}
\end{figure}

The point of the Imura's analysis seems to be nice fit of Eq.~(\ref{distribution}) with their distribution function. The function is obtained by avaraging over $5000$ samples and is used for intuitive understanding of the $S_{\chi}$ behavior, although his $S_{\chi}$ itself would be obtained from single snapshot. Here, it should be noted that we have two power-law functions: one is $f(\lambda)$, and the other one is $\lambda_{n}$ itself. Actually, as shown in Fig.~\ref{fig1}(a), we clearly see that $\lambda_{n}$ decays algebraically. Hence, in the scaling regime, it is possible for single snapshot data to denote
\begin{eqnarray}
\lambda_{n}=an^{-\Delta}, \label{lambda}
\end{eqnarray}
where $\Delta$ is a critical exponent and $a$ is a normalization constant. Then, we would like to consider how to relate $\Delta$ with $\alpha$ (or $\eta$) and whether this relation is really consistent with Fig.~\ref{fig1}(a).

For this purpose, we integrate Eq.~(\ref{distribution}) from $\lambda_{j}$ to $\lambda_{k}$ (note that $\lambda_{j}<\lambda_{k}$ for $j>k$)
\begin{eqnarray}
j-k = A\int_{\lambda_{j}}^{\lambda_{k}}d\lambda\lambda^{-\alpha} = \frac{A}{1-\alpha}\left(\lambda_{k}^{1-\alpha}-\lambda_{j}^{1-\alpha}\right).
\end{eqnarray}
Then, we have
\begin{eqnarray}
k=\frac{A}{\alpha-1}\left(ak^{-\Delta}\right)^{1-\alpha},
\end{eqnarray}
and we obtain
\begin{eqnarray}
\Delta=\frac{1}{\alpha-1}=1-\eta. \label{delta}
\end{eqnarray}
In the Ising model case, this is equal to $0.75$.

Let us go back to Fig.~\ref{fig1}(a), in which we show three spectra near $T_{c}=2.269J$ (the scaling regime for our finite size cluster is bit higher than this temperature). We find that all the data are consistent with Eqs.~(\ref{lambda}) and (\ref{delta}) in intermediate-$n$ region. Particularly, the spectrum at $T=2.35J$ fits very well with the scaling equation in wide-$n$ region, and the corresponding snapshot shown in Fig.~\ref{fig1}(c) is fractal-like spin structure with respect to critical fluctuation. Thus, by looking at small-$n$ region of the spectra, it is possible to find out how close our system approaches the critical regime.

Based on the eigenvalue spectrum, let us derive $S_{\chi}$ from Eq.~(\ref{lambda}). In order to determine $a$, we first represent the normalization condition for $\lambda_{n}$ as
\begin{eqnarray}
1=\sum_{n=1}^{L}\lambda_{n}\simeq\sum_{n=1}^{N}an^{-\Delta},
\end{eqnarray}
where $N$ represents the position where the numerical data start to deviate from the scaling line in large-$n$ region. In Fig.~\ref{fig1}, $N$ is roughly estimated to be $\ln N\simeq 5$. In the large-$L$ limit and at $T_{c}$, we expect that $N=L$. In this limit, the sum of $\lambda_{n}$ is equal to the zeta function, and the condition $\Delta=1-\eta<1$ means that the sum diverges. Thus $a$ should be a function of $L$, and become zero in this limit. In the continuous representation, we have
\begin{eqnarray}
1\simeq a\int_{1}^{N}dx x^{-\Delta} = \frac{a}{1-\Delta}\left(N^{1-\Delta}-1\right).
\end{eqnarray}
and then $a$ is given by
\begin{eqnarray}
a=\frac{1-\Delta}{N^{1-\Delta}-1}.
\end{eqnarray}
Next, the entropy is evaluated as
\begin{eqnarray}
S_{\chi} &\simeq& -\int_{1}^{\chi}dx\left(ax^{-\Delta}\right)\ln\left(ax^{-\Delta}\right) \nonumber \\
&=& - \frac{a}{1-\Delta}\left( \ln a + \frac{\Delta}{1-\Delta} \right)\left(\chi^{1-\Delta}-1\right) \nonumber \\
&& + \frac{a\Delta}{1-\Delta}\chi^{1-\Delta}\ln\chi \nonumber \\
&=& \frac{\chi^{1-\Delta}-1}{N^{1-\Delta}-1}\left\{\ln\left(N^{1-\Delta}-1\right)-\gamma(\Delta)\right\} \nonumber \\
&& + \frac{\Delta}{N^{1-\Delta}-1}\chi^{1-\Delta}\ln\chi, \label{schi}
\end{eqnarray}
with
\begin{eqnarray}
\gamma(\Delta)=\ln(1-\Delta)+\frac{\Delta}{1-\Delta}. \label{gamma}
\end{eqnarray}
We find that the last term in Eq.~(\ref{schi}) is the leading one and is consistent with Imura's result.

An advantage of the present method is that we can also obtain the full snapshot entropy. When we take $N=\chi=L\gg 1$, we obtain
\begin{eqnarray}
S_{L} = \ln L - \gamma(\Delta).
\end{eqnarray}
In this limit, the power factor vanishes due to the presence of $N^{1-\Delta}$. In the Ising model case, the $\gamma$ value is taken to be $\gamma(1-\eta) = 3 - 2\ln 2 = 1.6137$. This is somehow smaller than the estimated value in Ref.~\cite{Matsueda}. The $\eta$ dependence of $\gamma$ in Eq.~(\ref{gamma}) is remarkable in small-$\eta$ region, and thus we may need more sophisticated calculation to strictly determine the $\gamma$ value in the numerical side.

Summarizing, we basically agree with Imura et al., but at the same time we emphasize that the power factor $\chi^{\eta}$ vanishes in the full entropy. Thus the Calabrese-Cardy type scaling without the anomalous dimension is robust. As we have discussed in Ref.~\cite{Matsueda2}, the change in the finite-$\chi$ scaling is remarkable, when the full conformal symmetry breaks on the image.

\end{document}